\documentclass{article}
\usepackage{latexsym,amsmath,amsopn,amssymb,amsthm,amsfonts}
\usepackage{graphicx} 
\usepackage{authblk}
\usepackage[T2A]{fontenc}
\usepackage[cp1251]{inputenc}
\usepackage{amsmath}
\usepackage[mathscr]{eucal}
\usepackage[english]{babel}
\usepackage[left=2.4cm, right=2cm, top=2cm, bottom=2.4cm]{geometry}

\title{Noether's theorem  \\ for the conditional principle of least action}
\author[1]{S.L.~Lyakhovich}
\author[1]{S.B.~Sayapin}
\author[2]{I.A.~Zubareva}

\affil[1]{\small{Tomsk State University, Novosobornaya Sq 1, Tomsk 634050, Russia, }\protect\\
E-mail: \texttt{sll@phys.tsu.ru}, \texttt{s.sayahin@phys.tsu.ru}}
\affil[2]{\small{Omsk Department of Sobolev Institute of Mathematics, Pevtsova St 13, Omsk 644099, Russia,}\protect\\ 
E-mail: \texttt{izubareva@ofim.oscsbras.ru}}

\date{}

\begin{document}

\maketitle
\begin{abstract}
 We consider the problem of a conditional extremum of an action in a class of fields constrained by differential equations. For this setup, we propose an extension of Noether's first theorem to connect the symmetries of the action and the imposed equations to the currents conserved at the conditional extrema.
 The key ingredient of the extension is the gauge symmetry of the differential equations constraining the admissible class of field configurations. 
 We consider a special type of global symmetries of the action which we call conditional symmetries. Such global symmetries must be special cases of gauge transformations of the constraint equations. We construct conservation laws that follow from the conditional symmetries of the action. No Lagrange multipliers or other auxiliary fields are introduced and the conserved currents include only the original fields. We also prove the converse theorem which connects the conserved currents to the conditional symmetries of the action.
The general method is illustrated by several examples.
\end{abstract}
\maketitle

\section{Introduction}
We consider the problem of the conditional principle of least action, where the class of admissible field configurations is restricted by differential equations.
It is known that the inclusion of constraints with Lagrange multipliers provides the necessary conditions for the conditional extremum of the functional, but the multipliers can introduce additional degrees of freedom that are not related to the original dynamics\footnote{For holonomic constraints, Lagrange multipliers do not introduce any additional degree of freedom and can be algebraically eliminated from the equations. Another notable exception: equations in which all degrees of freedom are gauged out when  the constraints are considered without any connection to the action. Including these equations with Lagrange multipliers in the action does not introduce new degrees of freedom. For general differential equations, Lagrange multipliers create new degrees of freedom.}. The conserved Noether currents for the action with Lagrange multipliers will involve these non-physical degrees of freedom. So, these conserved currents have no direct interpretation from the perspective of the original conditional extre\-mum problem. We propose a different idea for constructing conserved quantities related to global symmetries of the conditional extremum problem, so that the resulting currents involve only the original degrees of freedom. 

Let us give a preliminary general explanation of the main idea of the article, which is implemented in four stages.  At first, we should find the infinitesimal gauge symmetry transformations of the equations restricting the fields. The second step is to derive the field equations by imposing the requirement that the gauge variation of the action must vanish for arbitrary gauge parameters. These equations must be satisfied for the action to have a conditional extremum, as shown in the article \cite{Lyakhovich:2025}. Third, it is necessary to find the global symmetries of the action being a specialization of the gauge symmetry transformations of the equations that restrict the admissible class of trajectories. By specialization we mean gauge transformations with specific parameters that involve only arbitrary constants, not arbitrary functions. We call these residual gauge transformations conditional symmetries of the action. At the fourth stage, we derive conserved currents that correspond to the conditional symmetries of the action. The inverse mapping, from conserved currents to conditional symmetries of the action, is obtained by reversing the same procedure.

Given the vast body of literature on Noether's theorems, a general overview of this topic is beyond the scope of this article, which deals specifically with the problem of conditional extremum. For a review of the general area of Noether's theorem and a detailed bibliography we refer to the monograph \cite{{Kosmann}}. We can also note that if the field equations do not follow from the unconditional variational principle but admit some additional structures, then a one-way relationship between global symmetries and conservation laws can be possible. For example, a pre-symplectic structure \cite{Crnkovic}, \cite{Havkine}  maps symmetries to conserved currents, \cite{Grigoriev-presymplectic}, \cite{Grigoriev-presymplectic2}, \cite{Sharapov-tricomplex}, \cite{Sharapov-presymplectic}, and a Lagrange anchor \cite{Kazinski:2005eb} defines an inverse mapping from conservation laws to symmetries \cite{Kaparulin:2010ab}. In this paper, we construct a generalization of the first Noether theorem, and the converse one, for the case of the conditional extremum problem. We proceed from the usual definitions and assumptions related to the issues of symmetries and conserved currents, known from various textbooks on field theory, see, for example, \cite{bogolubov-shirkov}, while for the gauge symmetry we use the standard setup and notation of gauge theories \cite{Henneaux:1992ig}.   

The article is organised as follows. In the next section, we briefly outline the conditional Lagrange equations that follow from the conditional principle of least action. The conditional extremum of the action is sought in the class of admissible fields that obey differential equations. This Section also addresses the gauge symmetry of the equations that constrain the fields. In Section 3, we consider special infinitesimal global symmetries of the action in the class of the fields subject to differential equations. These conditional symmetries of the action represent specializations of gauge transformations of the equations that are imposed to constrain the admissible class of fields. We then show that such conditional global symmetries lead to currents that are conserved at conditional extrema of the action. This fact is a main result of the article. We also prove the converse theorem: Every current, which is conserved at conditional extremals of the action, leads to a conditional symmetry. Throughout this paper, we develop the formalism in its full generality, without restricting the order of derivatives. At the end of Section 3, we provide a brief exposition of the key relations, including an explicit construction of conserved currents for the simplest case where the Lagrangian does not include higher-order derivatives of the fields, and gauge transformations of the constraint equations do not involve higher-order derivatives of the gauge parameters.
In Section 4, we illustrate the general formalism of Sections 2-3 by examples. The final section discusses the results and possible areas of application of the proposed formalism, and more broadly, of conditionally Lagrangian theories in general.

\section{Field equations for the problem of conditional extremum}
Consider a set of $n$ fields $\phi^i(x),\, i=1,\ldots , n$. In the simplest case, the fields can be thought of as real smooth functions of the space-time coordinates $x^\mu, \mu=0,1,\ldots,d-1$.
The principle of conditional least action assumes that true configurations are defined as action extremals in a class of fields that obey a system of differential equations. 

Consider the action functional of the usual form 
\begin{equation}\label{S}
S[\phi(x)]=\int d^d x L(x,\phi, \partial\phi,\partial^2\phi,\ldots)
\end{equation}
and impose the differential equations to restrict the admissible fields,
\begin{equation}\label{T}
T_a(x,\phi, \partial\phi, \partial^2\phi,\ldots)=0,  \quad  a=1,\ldots ,N \, .
\end{equation}
The problem is to find the minimum of the action (\ref{S}) among the fields subject to equations (\ref{T}).

From the perspective of the calculus of variations, there are two main ways to solve the problem of conditional extremum \cite{Gelfand}. 
First, the constraints (\ref{T}) with Lagrange multipliers can be added to the action. This reduces the problem to the issue of unconditional extremum for the action with Lagrange multipliers
\begin{equation}\label{S-lambda}
 S_\lambda[\phi(x),\lambda(x)]=\int dx \left( L(x,\phi, \partial\phi,\partial^2\phi,\ldots) + \lambda^a T_a (x,\phi, \partial\phi, \partial^2\phi,\ldots) \right) \, .
\end{equation}
Corresponding  Euler-Lagrange equations 
\begin{equation}\label{Eq-S-lambda}
  \frac{\delta S_\lambda}{\delta \phi^i}=0\, , \quad    \frac{\delta S_\lambda}{\delta \lambda^a}\equiv T_a (x,\phi, \partial\phi, \partial^2\phi,\ldots)=0\, 
\end{equation}
provide necessary conditions for the extremum of the original functional (\ref{S}) in the class of fields restricted by equations (\ref{T}).
The Lagrange multipliers, being involved in these equations, can bring, in general, extra degrees of freedom comparing to the original problem of conditional extremum if equations (\ref{T}) are differential. If the Noether theorem is applied immediately to the action (\ref{S-lambda}), corresponding conserved currents will involve Lagrange multipliers, and the non-physical degrees of freedom will contribute to the conserved quantities. These conserved quantities seem having no direct connection with the original problem of conditional ext\-re\-mum\footnote{In special cases, already mentioned in the introduction, including constraints with Lagrangian multipliers in the action do not introduce additional degrees of freedom into the theory. In such cases, the conserved currents for the action with multipliers (\ref{S-lambda}) will depend on these multipliers but will not involve additional degrees of freedom. In such special cases, the multipliers can be eliminated on the mass shell, and the resulting currents will include only the degrees of freedom of the original conditional extremum problem.}.  

The second way to solve the conditional extremum problem is to use unfree variations of the fields $\delta'\phi^i$. These variations must respect equations (\ref{T}),
\begin{equation}\label{delta-prime}
  \delta' T_a{}_{\big|T=0}=0 \, .
\end{equation}
The above equations are the restrictions imposed on the unfree variations of the fields. Explicitly, these restrictions read
\begin{equation}\label{delta-prime-T}
  \sum_{k=0}\left(\frac{\partial T_a(\phi,\partial\phi,\partial^2\phi, \dots )}{\partial\, \partial_{\mu_1}\ldots\partial_{\mu_k} \phi^i}\right)_{\big|T=0} \partial_{\mu_1}\ldots\partial_{\mu_k}\delta'\phi^i = 0\, .
\end{equation}
These relations are the linear differential equations for variations $\delta'\phi^i$. The coefficients of these linear equations depend on the fields $\phi^i$. The latter variables are assumed to obey equations (\ref{T}). 
For the action (\ref{S}) to have an extremum under the constraints (\ref{T}) imposed on the fields, it is necessary that the unfree variation of $S$ vanishes for any variations  $\delta'\phi$ consistent with (\ref{delta-prime-T}),
\begin{equation}\label{delta-prime-S}
\delta' S=\int dx \left(\frac{\delta S}{\delta\phi^i}\delta'\phi^i  + total \,\, divergence\, \right) = 0,
\end{equation}
and for field configurations admitted by the constraints (\ref{T}).
The above relation does not immediately lead to a system of differential equations, unlike the variational principle for the action with Lagrange multipliers (\ref{Eq-S-lambda}), because the variations $\delta'\phi^i$, being restricted by equations  (\ref{delta-prime-T}), are not arbitrary functions.
As a result, the infinitesimal symmetries of the action, even if equations (\ref{T}) have the same symmetry, do not in general lead to currents that are conserved at the conditional extrema of $S$.

Now let us briefly explain, following \cite{Lyakhovich:2025}, how to explicitly construct the field equations that are necessary conditions for the action extremum in the class of fields constrained by differential equations. 
First, we should find the infinitesimal gauge symmetry transformations for equations (\ref{T}) which are imposed to select the admissible field configurations. It is assumed that the transformation includes derivatives of the gauge parameters $\epsilon^\alpha(x), \alpha=1,\ldots,m$, which are arbitrary smooth functions, up to a certain finite order $K$ 
 \begin{equation}\label{GT}
\delta_\epsilon\phi^i=\hat{R}^i_\alpha\epsilon^\alpha=\sum_{k=0}^{K}\stackrel{(k)}{R}{}^{i}_\alpha{}^{\mu_1\ldots\mu_k} (x,\phi,\partial\phi,\partial^2\phi\ldots)
\partial_{\mu_1}\ldots\partial_{\mu_k}\epsilon^\alpha\, ,
\end{equation}
where $\stackrel{(k)}{R}{}^i_\alpha{}^{\mu_1\ldots\mu_k}$ are smooth functions of the fields and their derivatives.
The above transformation is supposed to be a symmetry of equations (\ref{T}), hence
\begin{equation}\label{delta-epsilon-T}
 \delta_\epsilon T_a{}_{\big|T=0}=0, \quad \forall \epsilon_\alpha (x) \, .
\end{equation}
Under appropriate assumptions on regularity for the functions $T_a$, this means
\begin{equation}\label{U}
 \delta_\epsilon T_a= \hat{U}^b_a T_b  \, ,
\end{equation}
 where $\hat{U}^b_a$ is a differential operator whose coefficients depend on the fields, gauge parameters and their derivatives.
It is assumed that the generating set of gauge transformations $\hat{R}^i_\alpha\epsilon^\alpha$ is (over-)complete in the sense that any gauge transformation that preserves equations (\ref{T}), i.e. has the property (\ref{delta-epsilon-T}), is spanned by transformations (\ref{GT}).

Given any variation of the fields, the variation of the action reads 
\begin{equation}\label{delta-S}
\delta S[\phi(x)]=\int dx\left(\partial_\mu J^\mu+\frac{\delta S}{\delta\phi^i}\delta\phi^i\right)\, ,
\end{equation}
\begin{equation}\label{J}
J^\mu=\sum_{p=1}^{M} p^{\mu \mu_1\ldots\mu_{p-1}}_i\partial_{\mu_1}\ldots\partial_{\mu_{p-1}}\delta \phi^i \, ,
\end{equation}
 \begin{equation}\label{p}
 p^{\mu \mu_1\ldots\mu_{p-1}}_i=\sum_{k=0}^{M-p}(-1)^k\partial_{\nu_1}\ldots\partial_{\nu_k}
 \frac{\partial L}{\partial(\partial_{\nu_1}\ldots\partial_{\nu_k}\partial_{\mu}\partial_{\mu_1}\partial_{\mu_{p-1}}\phi^i)}\, ,
 \end{equation}
 where $M$ is the maximal order of the derivatives in  the Lagrangian.
 The variation must vanish at a conditional extremum of the action for any variations of the fields that are consistent with equations (\ref{T}) in the sense of the relations (\ref{delta-prime}). The requirement  (\ref{delta-prime-S})  in itself does not immediately lead to equations that determine the conditional extremals, since the variations of the fields are not free, but are restricted by the differential equations (\ref{delta-prime-T}).
 
Substituting the gauge variation of the fields (\ref{GT}) into the general variation of the action (\ref{delta-S}) and integrating by parts we come to the following expression 
\begin{equation}\label{delta-epsilon-S}
  \delta_\epsilon S  =  \int dx\left( \epsilon^\alpha\hat{R}^{\dagger i}_\alpha \, \frac{\delta S}{\delta\phi^i} + 
  \partial_\mu I^\mu\right)   \, ,
\end{equation}
where $\hat{R}^{\dagger i}_\alpha$ is  the formal Hermitian conjugate operator for the gauge generator $\hat{R}^i_\alpha$ (\ref{GT}),
\begin{equation}\label{R-dagger}
\hat{R}^{\dagger i}_\alpha F =\sum_{k=0}^{K}(-1)^k\partial_{\mu_1}\ldots\partial_{\mu_k}\left(\stackrel{(k)}{R}{}^{i\mu_1\ldots\mu_k}_\alpha F\right),
\end{equation}
and the current $J^\mu$ (\ref{delta-S}), (\ref{J}) changes to an additional term
\begin{equation}\label{I}
I^\mu=J^\mu+\sum_{p=0}^{K-1}\left(\left(\partial_{\mu_1}\ldots\partial_{\mu_p}\epsilon^\alpha\right)\sum_{k=0}^{K-p-1}\left(\partial_{\mu_{p+1}}\ldots\partial_{\mu_{p+k}}\left(\frac{\delta S}{\delta\phi^i}(-1)^k\stackrel{(k+p+1)}{R}{}^i_\alpha{}^{\mu\mu_1\ldots\mu_{p+k}}\right)\right)\right) \, .
\end{equation}
This additional term can contribute to the conserved currents for the problem of conditional extremum as we will see in the next section.

\noindent By construction, the gauge transformation (\ref{GT}), being consistent with equations (\ref{T}), does not take fields out of the admissible class (\ref{delta-epsilon-T}), while the gauge parameters are unrestricted arbitrary functions. Hence, the gauge variation (\ref{delta-epsilon-S}) must necessarily vanish at any conditional extremum of the action and for arbitrary gauge parameters $\epsilon^\alpha (x)$. This leads to the differential equations being necessary for the conditional extremum of the action
\begin{equation}\label{Partially-Lagrangian}
\frac{\delta_\epsilon S}{\delta\epsilon^\alpha}=\hat{R}^{\dagger i}_\alpha\frac{\delta S}{\delta\phi^i}=0\, ,\qquad T_a(x,\phi, \partial\phi, \partial^2\phi,\ldots)=0\, .
\end{equation}
We call the above system  \emph{conditionally Lagrangian equations}\footnote{Equations (\ref{Partially-Lagrangian}) imposed to define the fields being the conditional extremals of the action, were first proposed in the article \cite{Lyakhovich:2025}. In this article, equations (\ref{Partially-Lagrangian}) are referred to as a partially Lagrangian system. Here, we change the terminology to better suit the problem setup.}. 

It is worth noting that any configuration of the original fields $\phi^i$ determined by the Euler-Lagrange equations for the action with Lagrange multipliers (\ref{Eq-S-lambda}) will also satisfy the conditionally Lagrangian equations (\ref{Partially-Lagrangian}). 
This fact is proved in Ref. \cite{Lyakhovich:2025}.   This does not imply equivalence between the equations (\ref{Partially-Lagrangian}) and (\ref{Eq-S-lambda}), 
since the latter system includes Lagrange multipliers. 
The multipliers may require, in general, additional initial data. 
This introduces an additional degree of freedom that is not related to the conditional extremum problem as such. 
The equivalence of these systems is realized only in some special cases, including the case of holonomic constraints $T_a$ and some special cases of differential equations (\ref{T}), as explained in Ref. \cite{Lyakhovich:2025}.  
Both systems of equations, (\ref{Eq-S-lambda}) and (\ref{Partially-Lagrangian}), provide the necessary conditions for the conditional extremum of the action. 
However, system (\ref{Partially-Lagrangian}), which does not involve Lagrangian multipliers, is stronger  in the sense that it has less degree of freedom, in general. 
This difference in degrees of freedom can be seen in the examples in Section 4.

Let us finalise this section with remarks on the possible overlap between gauge symmetry of equations (\ref{T}) constraining the fields and the symmetry of the action.
If the gauge symmetry of the action includes all the gauge symmetry of the constraints, and maybe more, then the conditionally Lagrangian equations will reduce to constraints (\ref{T}) and gauge identities. In this case the dynamics will be totally defined by the constraints while minimisation of the action will not further restrict the fields. The most extreme case of this phenomenon is the system of constraints (\ref{T}) without any gauge symmetry. For the conditional Lagrangian equations to be truly non-trivial, the gauge symmetry of the action must be lower than the symmetry of the constraints restricting the admissible fields.
If the action has a gauge symmetry that is part of the gauge symmetry of equations (\ref{T}), this will lead to the Noether identities between the conditionally Lagrangian equations (\ref{Partially-Lagrangian}). The corresponding modification of Noether's second theorem and the related Batalin-Vilkovisky master equation will be addressed elsewhere.
Here we focus on the case where some gauge symmetries of the constraints (\ref{T}) do not leave the action invariant when the gauge parameters are arbitrary functions, but for certain constant gauge parameters these transformations preserve the Lagrangian modulo the total divergence. In the next section, we will show that symmetries of this type lead to currents that are conserved by virtue of the conditionally Lagrangian equations.

\section{Nother's theorem for the problem of conditional extremum}
Let us consider gauge symmetry transformations (\ref{GT}) of equations (\ref{T}) restricting the class of admissible fields, and choose the gauge parameters $\epsilon^\alpha (x)$ in the special way, 
\begin{equation}\label{omega}
\epsilon^\alpha_\omega(x)=\psi^\alpha_A(x, \phi,\partial\phi,\partial^2\phi,\ldots)\,\omega^A\, , \quad A=1,\dots, s \, ,
\end{equation}
where  $\psi^\alpha_A(x, \phi,\partial\phi,\partial^2\phi,\ldots)$ are specific functions of the fields and their derivatives, and  $\omega^A$ are arbitrary constant parameters. 
Since the original gauge transformations (\ref{GT}) are para\-me\-te\-rised by arbitrary functions, it is an infinite dimensional symmetry. 
Special gauge transformations, being para\-me\-te\-rised by constants (\ref{omega}), are a finite subset of this infinite symmetry.
We refer to the corresponding field transformations
\begin{equation}\label{delta-omega}
\tilde{\delta}_\omega \phi^i = \delta_\epsilon\phi^i_{|\epsilon\,\mapsto\,\epsilon_\omega}
\end{equation}
as the specializations of gauge transformations. Technically, specialization means that the gauge parameters are chosen as a linear combination of arbitrary constants, with the coefficients being certain functions of the fields and their derivatives.

\textbf{Definition.} \emph{A specialization of the gauge symmetry transformations of the equations that restrict the class of admissible fields is called  \textbf{a conditional symmetry of the action} if it leaves the Lagrangian invariant modulo the total divergence for the admissible fields.}

Given gauge symmetry (\ref{GT}) and the specialisation (\ref{omega}), (\ref{delta-omega}), the conditional symmetry of the action is defined by the relation
\begin{equation}\label{delta-omega-L}
\tilde{\delta}_\omega L = \omega^A\left(\partial_\mu\Lambda_A^\mu(\phi,\partial\phi, \dots)+\hat{W}^a_A T_a\right), 
\end{equation}  
where $\hat{W}^a_A$ is a differential operator whose coefficients depend on the fields and their derivatives.  
The functions $\psi^\alpha_A$ that define the corresponding specialization of gauge transformations (\ref{omega}) are called as the generators of a conditional symmetry of the action.

Let us mention that not every infinitesimal symmetry shared by the action (\ref{S}) and equations (\ref{T}) meets the above definition, given the requirements (\ref{omega}), (\ref{delta-omega}). 
\vspace{0.2 cm}

\noindent
\textbf{Theorem 1.}
For every conditional symmetry of a system whose equations of motion follow from the conditional principle of least action, there is a corresponding conserved quantity.

\vspace{0.2 cm}

\noindent
\textbf{Proof:}
Under the general gauge symmetry transformation (\ref{GT}), the Lagrangian transforms by the rule (\ref{delta-epsilon-S}). Substituting the specialization (\ref{omega}), (\ref{delta-omega}) into the gauge transformation of action (\ref{delta-epsilon-S}), we arrive at the relation
\begin{equation}\label{delta-L-I-omega}
 \tilde{\delta}_\omega L = \omega^A\left(\partial_\mu I^\mu_A +\psi^\alpha_A\hat{R}^{\dagger i}_\alpha\frac{\delta S}{\delta\phi^i} \right) \, ,    
\end{equation}
where currents $I_A^\mu$ read 
\begin{equation}\label{I-A}
\begin{split}
&\hspace{4cm} I_A^\mu=\sum_{p=1}^{M} p^{\mu \mu_1\ldots\mu_{p-1}}_i\partial_{\mu_1}\ldots\partial_{\mu_{p-1}}\hat{R}^i_\alpha\psi^\alpha_A\,+ \\
&\sum_{p=0}^{K-1}\left(\left(\partial_{\mu_1}\ldots\partial_{\mu_p} \psi^\alpha_A\right)\sum_{k=0}^{K-p-1}\left(\partial_{\mu_{p+1}}\ldots\partial_{\mu_{p+k}}\left(\frac{\delta S}{\delta\phi^i}(-1)^k\stackrel{(k+p+1)}{R}{}^{i}_\alpha{}^{\mu\mu_1\ldots\mu_p\mu_{p+1}\ldots\mu_{p+k}}\right)\right)\right) \, ,
\end{split}
\end{equation}
  and the quantities $p^{\mu \mu_1\ldots\mu_{p-1}}_i$ are defined by relations (\ref{p}).
  
Since the transformation $\tilde{\delta}_\omega\phi^i$ is assumed to be a conditional symmetry of the action, the corres\-pon\-ding variation of the Lagrangian (\ref{delta-L-I-omega}) must reduce to the total divergence (\ref{delta-omega-L}). As the parameters $\omega^A$ are arbitrary constants, this leads to identical equality:
\begin{equation}\label{tilde-I}
  \partial_\mu\tilde{I}{}^\mu_A=-\psi^\alpha_A\hat{R}{}^{\dagger i}_\alpha\frac{\delta S}{\delta \phi^i}+\hat{W}^a_AT_a \, , \quad \tilde{I}{}^\mu_A= I^\mu_A- \Lambda^\mu_A \, .
\end{equation}
This off-shell identity means that the currents $\tilde{I}{}^\mu_A$  are conserved by virtue of equations (\ref{Partially-Lagrangian}).

\vspace{0.1 cm}
\noindent \textbf{Q.E.D.}

\vspace{0.1 cm}

Let us now formulate and prove the converse statement.
\vspace{0.1 cm}

\noindent
\textbf{Theorem 2.}
For every conserved current of the system whose equations of motion follow from the conditional principle of least action, there is a corresponding conditional symmetry of the action.

\vspace{0.1 cm}

\noindent
\textbf{Proof:} Given the set of currents $\Phi^\mu_A$ which are conserved by virtue of the field equations
(\ref{Partially-Lagrangian}), they must satisfy the following off-shell identity
\begin{equation}\label{characteristics}
  \partial_\mu{\Phi}{}^\mu_A=\hat{\Psi}{}^\alpha_A\hat{R}{}^{\dagger i}_\alpha\frac{\delta S}{\delta \phi^i}+\hat{W}^a_AT_a \, , 
\end{equation}
where $\hat{\Psi}{}^\alpha_A, \, \hat{W}^a_A $ are differential operators\footnote{Under standard regularity assumptions, if a smooth function (not necessarily a total divergence) of the fields and their derivatives vanishes on-shell, then it can be expressed as a linear combination, with coefficients that may depend on the fields and their derivatives, of the left-hand sides of the field equations and their differential consequences, \cite{Henneaux:1992ig}.}.
The coefficient $\hat{\Psi}{}^\alpha_A$ is called the conditional characteristics of the conserved current. It is assumed non-vanishing at the general admissible field configurations,
\begin{equation}\label{Psi-hat}
 \hat{\Psi}^\alpha_A=\sum_{k=0}^{N}\stackrel{(k)}{\psi}{}^{\alpha \, \mu_1\ldots\mu_k}_A(\phi,\partial\phi,\partial^2\phi,\ldots)\partial_{\mu_1}\ldots\partial_{\mu_k} \, ,\qquad
 \hat{\Psi}{}^\alpha_A{}_{\big|T=0}\neq 0, \quad\forall A \, .
\end{equation}
Conserved current $\Phi{}^\mu_A$, being defined by relations (\ref{characteristics}), can be re-defined by adding on shell vanishing terms to exclude the derivatives from the characteristics (\ref{Psi-hat}),
\begin{equation}\label{psi}
\partial_\mu\Phi^{\prime\mu}_A=\hat{W}^a_AT_a+\psi^\alpha_A\hat{R}^{\dagger i}_\alpha\frac{\delta S}{\delta\phi^i},
\end{equation}
where
\begin{equation}\label{Phi-prime}
\resizebox{0.9\textwidth}{!}{$\displaystyle
\Phi^{\prime\mu}_A=\Phi^\mu_A-\sum_{p=0}^{N-1}\bigg(\left(\partial_{\mu_1}\ldots\partial_{\mu_p}\hat{R}^{\dagger i}_\alpha\frac{\delta S}{\delta\phi^i}\right)
\sum_{k=0}^{N-p-1}\bigg(\partial_{\mu_{p+1}}\ldots\partial_{\mu_{p+k}}\bigg((-1)^k\stackrel{(k+p+1)}{\psi}{}^{\alpha}_A{}^{\mu\mu_1\ldots\mu_p\mu_{p+1}\ldots\mu_{p+k}}\bigg)\bigg)\bigg) \, ,$}
\end{equation}
\begin{equation}\label{Psi-psi}
\psi^\alpha_A=\sum_{k=0}^{N}(-1)^k\partial_{\mu_1}\ldots\partial_{\mu_k}\stackrel{(k)}{\psi}{}^{\alpha}_A{}^{\mu_1\ldots\mu_k}\, ,
\end{equation}
and $\psi^\alpha_A$ are the functions of the fields and their derivatives. As we see from (\ref{psi}), the conditional characteristics of the conserved current can be taken to be a set of functions, not differential operators. Let us also mention that the conditional characteristics $\psi^\alpha_A$ is assumed non-vanishing at general admissible fields defined by equations (\ref{T}). If $\psi^\alpha_A=0$ for any solution of equations (\ref{T}), there still can be a nontrivial conserved current, as we see from (\ref{psi}). This current will be conserved only by virtue of equations (\ref{T}) that restrict the admissible field configurations, and regardless of the action. This conservation law is obviously unrelated to any symmetry of the action. For example, equations (\ref{T}) can be conservation laws by themselves. Obviously, if this conservation law is imposed to select the admissible fields, it is automatically consistent with the conditionally Lagrangian equations (\ref{Partially-Lagrangian}), irrespectively to any symmetry of the action. To connect the conserved current with the symmetry, the conditional characteristics $\psi^\alpha_A$ should not vanish on shell.

Given the conserved current $\Phi^{\prime\mu}_A$ and corresponding conditional characteristics  $\psi^\alpha_A$ (\ref{psi}), let us construct a conditional symmetry of the action.
Consider a specialization of the gauge symmetry defined by the conditional characteristics (\ref{psi})  according to the rule (\ref{omega}), (\ref{delta-omega})
\begin{equation}\label{psi-omega}
\delta_\epsilon\phi^i=\hat{R}^i_\alpha \epsilon^\alpha, \quad  \epsilon_\omega = \psi^\alpha_A\omega^A\, ,\quad \tilde\delta_\omega\phi^i=\delta_{\epsilon_\omega}\phi^i \, .
\end{equation} 
Substituting the above specialization of the gauge transformation into the corresponding variation of Lagrangian (\ref{delta-L-I-omega}) and accounting for the conservation law (\ref{psi}), (\ref{Phi-prime}), we arrive at the conclusion
\begin{equation}
\tilde{\delta}_\omega L=\omega^A\left(\partial_\mu\left(I^\mu_A +\Phi^{\prime\mu}_A\right) +\hat{W}{}^a_A T_a \right)\, .
\end{equation}
As we see, the conditional characteristics defines a transformation (\ref{psi-omega}) that leaves the Lagrangian invariant modulo the total divergence for the admissible class of fields.

\noindent \textbf{Q.E.D.}

Theorem 1 can be viewed as an extension of Noether's first theorem to the case of a system whose field equations follow from the requirement of a conditional extremum of the action.
Theorem 2 is an extension of the converse statement to the same class of field theories.


\subsection*{Conditional symmetry and conserved currents without higher derivatives.}
The formalism considered in the two previous sections covers the general case where the Lagrangian can include derivatives of fields of any order, and the gauge transformations (\ref{GT}) can involve any derivatives of the gauge parameters. The full generality of the formalism may leave the impression that the construction of conditionally Lagrangian equations and the conserved currents is a rather cumbersome procedure.   However, Lagrangians without higher-order derivatives are most often studied, and gauge symmetries also most often include only the first derivatives of the parameters. In such a situation, without higher-order derivatives, both the conditional Lagrangian equations and the corresponding conditional symmetries and conserved currents take on a simpler form. Therefore, for the convenience of those readers who might be more interested in searching for conditional conservation laws in first-order models than in the generality of the formalism, we present a summary of the key expressions for this simpler type of theory. In this most common class of Lagrangians and gauge symmetries, the formalism is quite straightforward, and not much more complicated than the corresponding constructions in the unconditional action extremum problem.

Let us assume that gauge transformation (\ref{GT}) of the constraining equations (\ref{T}) involves at most the first derivatives of the gauge parameters,
\begin{equation}\label{GT1}
\delta_\epsilon\phi^i = \stackrel{(0)}{R}{}^i_\alpha (\phi,\partial\phi, \partial^2\phi, \dots)\epsilon^\alpha+\stackrel{(1)}{R}{}^i_\alpha{}^\mu (\phi,\partial\phi, \partial^2\phi, \dots) \partial_\mu\epsilon^\alpha   \, , \quad \delta_\epsilon T_a (\phi,\partial\phi, \partial^2\phi, \dots)_{\big|T=0} =0  \,, \,\, \forall\epsilon^\alpha (x) .
\end{equation}
Let us further assume that the Lagrangian does not involve higher derivatives. 
The conditional extrema of the action are subject to conditional Lagrangian equations (\ref{Partially-Lagrangian}), which in this case read:
\begin{equation}\label{CLE1}
\stackrel{(0)}{R}{}^i_\alpha \left(\frac{\partial L}{\partial\phi^i}-\partial_\mu\frac{\partial L}{\partial\partial_\mu\phi^i}\right) - \partial_\mu\left(\stackrel{(1)}{R}{}^i_\alpha{}^\mu \left(\frac{\partial L}{\partial\phi^i}-\partial_\mu\frac{\partial L}{\partial\partial_\mu\phi^i}\right)\right)=0\, ,\quad T_a (\phi,\partial\phi, \partial^2\phi, \dots) =0 \, .
\end{equation}
The specialization of gauge symmetry (\ref{GT1}) is constructed by choosing gauge parameters $\epsilon^\alpha (x)$ as the linear combinations of arbitrary constants $\omega^A$ with  field-dependent coefficients $\psi^\alpha_A (\phi,\partial\phi, \dots)$,
\begin{equation}\label{delta-omega-1}
\tilde{\delta}_\omega\phi^i = \left(\stackrel{(0)}{R}{}^i_\alpha \psi^\alpha_A +\stackrel{(1)}{R}{}^i_\alpha{}^\mu  \partial_\mu\psi^\alpha_A\right)\omega^A  \,  .
\end{equation}
This global transformation of fields is understood as a conditional symmetry of the action if it leaves the Lagrangian unchanged modulo the total divergence for constrained fields,
\begin{equation}\label{delta-omega-L1}
  \tilde{\delta}_\omega L_{\big| T=0}=\omega^A \partial_\mu \Lambda^\mu_A (\phi,\partial\phi, \dots) \, . 
\end{equation}
Given the above conditional symmetry, there is a current 
\begin{equation}\label{Current-1}
  \tilde{I}{}^\mu_A = \frac{\partial L}{\partial \partial_\mu \phi^i}\left(\stackrel{(0)}{R}{}^i_\alpha\psi^\alpha_A+\stackrel{(1)}{R}{}^i_\alpha{}^\nu\partial_\nu\psi^\alpha_A\right)+ \left(\frac{\partial L}{\partial\phi^i}-\partial_\nu\frac{\partial L}{\partial\partial_\nu\phi^i}\right) \stackrel{(1)}{R}{}^i_\alpha{}^\mu\psi^\alpha_A -\Lambda^\mu_A \, ,  
\end{equation}
 which is conserved by virtue of the conditional Lagrangian equations (\ref{CLE1}). With higher derivatives in gauge transformations and Lagrangian, this current would be defined by relations (\ref{I-A}),(\ref{tilde-I}).
 
 In this simplest case, the formalism of conditionally Lagrangian systems does not seem much more cumbersome compared to the usual construction of Noether currents for the unconditional extremum problem. 
 
 To conclude this subsection, let us relate these conditionally conserved currents, when the constraints are switched off, to the canonical Noether conservation laws for the unconditional extremum problem. Let no restrictions be imposed, then equations (\ref{T}) reduce to identities. The gauge symmetry of such trivial constraints is the shift of the fields by an arbitrary function $\delta_\epsilon \phi^i=\epsilon^i$. The complete set of gauge generators reads
  \begin{equation}\label{GT-T-0}
 \stackrel{(0)}{R}{}^i_j=\delta ^i_j\, , \quad  \stackrel{(1)}{R}{}^i_j{}^\mu =0 \, .
 \end{equation}
Clearly, any global symmetry of any action can be reproduced as a specialization of the above gauge transformation, since a shift of the field by an arbitrary function can absorb any infinitesimal specific transformation. For example, for Poincar\'e shift of Minkowski space by the constant vector $\omega^\mu$, $\delta_\omega\phi^i=\omega^\mu\partial_\mu\phi^i$ we have $\epsilon^i_\omega=\omega^\mu\partial_\mu\phi^i$. So, the generator of this ``conditional'' symmetry reads
  \begin{equation}\label{psi-1}
 \psi^i_\mu=\partial_\mu\phi^i \, .
 \end{equation}
Any  $x$-independent Lagrangian changes to a total divergence under this shift, 
$$\delta_\omega L(\phi,\partial\phi)=\omega^\mu\partial_\mu L=\partial_\mu (\Lambda^\mu_\nu \omega^\nu), \quad \Lambda^\mu_\nu=\delta^\mu_\nu L \, . $$
Substituting the above expressions for $R,\psi, \Lambda$ into the general expression for the conserved current (\ref{Current-1}), we arrive (up to an overall sign) to the canonical energy-momentum tensor,
\begin{equation}\label{TEM1}
  T^\mu_\nu =\frac{\partial L}{\partial\partial_\mu \phi^i}\partial_\nu\phi^i\,-\,L\delta^\mu_\nu \, .
\end{equation}
As can be clearly seen, once the equations (\ref{T}) restricting the fields disappear, the conditionally conserved currents (\ref{Current-1}) are reduced to the canonical conserved Noether quantities.

\section{Examples}
In this section we consider three model examples to illustrate the general formalism of the previous two sections. Two examples are mechanical systems, and one more is a field theory.  The actions are well known in all three cases, so one can see how restrictions imposed on the admissible configurations can significantly change the dynamics, including the conserved quantities.

\subsection*{Example 1. Particle in central field with conserved angular momentum.} 
Let us consider the planar motion of a point non-relativistic particle in a central potential 
\begin{equation}\label{S-particle}
S[r(t),\varphi(t)]=\int\limits_{t_1}^{t_2}\left(\frac{m\dot{r}^2}{2}+\frac{mr^2\dot{\varphi}^2}{2}-U(r)\right)dt,
\end{equation}
and  restrict the class of admissible trajectories by imposing the condition of conservation of angular momentum, 
\begin{equation}\label{M}
\frac{dM}{dt}=0,\quad M = mr^2\dot{\varphi} \, .
\end{equation}
If the restriction was not imposed, this equation would appear as a consequence of the Lagrangian equations for the same action. 
Since the class of admissible trajectories is now constrained from the outset by this equation, it includes paths that do not obey the Lagrange equations. This is why there may be non-Lagrangian critical lines for the conditional extremum problem.

According to the general procedure of Section 2, now we should find the gauge symmetry of equation (\ref{M}). It is a non-Lagrangian equation, so the Dirac-Bergmann algorithm does not work in this case. There is a general method of finding gauge symmetry for any ODE system \cite{Lyakhovich:2008hu}. 
This method leads to the following gauge symmetry transformation of equation (\ref{M}),     
\begin{equation}\label{GT-M}
\delta_\epsilon r=\dot{r}\epsilon-\frac{r}{2}\dot{\epsilon},\quad\delta_\epsilon \phi=\dot{\phi}\epsilon \, .
\end{equation}
One can verify by direct variation that $\delta_\epsilon \dot{M}=0,\, \forall\epsilon (t)$.
Given the gauge symmetry of the imposed constraint (\ref{M}), we arrive at equations (\ref{Partially-Lagrangian}) for the conditional extremals of the action (\ref{S-particle}) 
\begin{equation}\label{C-L-E-M}
\frac{\delta S}{\delta \epsilon}= \frac{d}{dt}\left(\frac{m\Ddot{r}r}{2}+\frac{m{\dot{r}}^2}{2}+U(r)+\frac{r}{2}\frac{d U}{d r}\right) =0\, , \quad \frac{dM}{dt}=0 \, .
\end{equation}
These equations have been derived for this example in the article \cite{Lyakhovich:2025}, where one can find their general solution. Below,
we consider the issue of connection between conditional symmetries of the system and conserved quantities.
The action (\ref{S-particle}) and the constraint (\ref{M}) are obviously invariant under rotations and time shifts,
\begin{equation}\label{omega-E1}
  \delta_1\phi=\omega_1\,, \,\, \delta_1r=0\, ; \qquad \delta_2 \phi=\dot{\phi}\,\omega_2\,,\,\, \, \delta_2 r =\dot{r}\,\omega_2 \, .
\end{equation}
In the case of an unconditional extremum, these symmetries immediately lead to the conservation laws of energy and angular momentum.
For a conditionally Lagrangian system, these symmetries must first be represented as a specialization of gauge transformations (\ref{GT-M}), and then conservation laws must be constructed by the general formulas (\ref{I-A}), (\ref{tilde-I}).
Comparing global symmetries (\ref{omega-E1}) with the gauge transformation  (\ref{GT-M}), we can immediately see the specialization for the time shift. It is sufficient just to take a constant gauge parameter:
\begin{equation}\label{T-shift-M}
  \epsilon(t)= \omega=const \, .
\end{equation}
In the case of rotation, the specialization of the gauge transformation is less immediately obvious,
\begin{equation}\label{rotation-M}
  \epsilon_\omega=r^2\omega, \quad \omega=const \, .
\end{equation}  
Given the constraint $\dot{M}=0$, this specialization of the gauge parameter leads to the transformation
\begin{equation}\label{rotation-M-dot}
\delta_\omega r=0,\quad \delta_\omega\phi=\frac{M}{m}\omega \, .
\end{equation}
\noindent For the admissible trajectories, where $M=const$, it is a rotation. 
Keeping in mind the imposed restriction $\dot{M}=0$, the angular velocity $\dot{\phi}$ is invariant under this transformation, while the Lagrangian does not involve $\phi$, so it is a conditional symmetry indeed. 
Substituting the generator of this conditional symmetry $\psi=r^2$ (see in (\ref{rotation-M})) into the general formula for conservation law (\ref{I-A}), (\ref{tilde-I}), we find the conserved quantity   
\begin{equation}\label{tilde-I-1-M}
\tilde{I}_1=\frac{M^2}{m}-4K,
\end{equation}
where 
\begin{equation}\label{K}
K=\frac{r^3}{8}\left(-m\ddot{r}-\frac{dU}{dr}+mr\dot{\phi}^2\right) \, .
\end{equation}
Since the angular momentum $M$ is conserved due to the constraint imposed from the outset, it is the quantity $K$ that is conserved as a consequence of isotropy. 
This conserved quantity was first noted in the article \cite{Lyakhovich:2025}, where it was called the constant of precession. In the case of the Kepler potential, this quantity is related to the precession velocity of bounded trajectories being solutions to conditionally Lagrangian equations (\ref{C-L-E-M}). For $K=0$ (this is an admissible initial condition) we arrive at the usual Lagrangian trajectories of a particle in the central potential.
 
Consider now the specialization (\ref{T-shift-M}) of the gauge symmetry (\ref{GT-M}) which defines the time shift. The generator is $\psi=1$,
\begin{equation}\label{time-shift-M} 
\delta_\epsilon r=\dot{r}\omega, \quad \delta_\epsilon\phi=\dot{\phi}\omega \, .
\end{equation}
Under this transformation,  the Lagrangian, being explicitly independent of time,  changes to a total derivative
\begin{equation}\label{delta-omega-L-M}
  \delta_\omega  L= \frac{d L}{dt}\omega\, .
\end{equation}
Substituting the conditional characteristics $\psi=1$ and the above total derivative ( cf. (\ref{delta-omega-L})) into general relations (\ref{I-A}), (\ref{tilde-I}) we arrive at the conserved quantity
\begin{equation}\label{E-M}
E=\frac{m\dot{r}^2}{2}+\frac{mr\ddot{r}}{2}+\frac{r}{2}\frac{dU}{dr}+U(r) \,  .
\end{equation}
Since the conservation of this quantity is associated with the homogeneity of time, it is the quantity that should be considered as the energy of the conditionally Lagrangian system (\ref{C-L-E-M}).
Making use of the conservation of the angular momentum $M$ (\ref{M}) and precession constant $K$ (\ref{K}), the above expression for energy can be reorganised. At first, we exclude $\ddot{r}$ using the conservation of $K$ (\ref{K}). This involves $\dot{\phi}$ which we exclude using the conservation of momentum $M$ (\ref{M}). As a result we arrive to the ordinary first order equation for $r$,
\begin{equation}\label{U-eff-M}
  E=\frac{m\dot{r}{}^2}{2}+U_{eff}(r)\, , \quad U_{eff}(r)=U(r) + \frac{M^2-8Km}{2mr^2} .
\end{equation}
From the point of view of the effective potential energy, the consequence of imposing the constraint on $M$ is a modification of the term with centrifugal energy, which now involves an additional constant of motion K. With $K=0$ (this is an admisible initial condition), we return to the usual Lagrangian dynamics. Given the conserved quantities $M,K,E$, the conditional Lagrangian equations (\ref{C-L-E-M}) are always reduced to quadratures using the effective potential (\ref{U-eff-M}), just as in the problem the of unconditional extremum.

Let us briefly comment on the nature of the dynamics for this particular conditional Lagrangian system for the case of the Kepler potential
 $U(r)=-\alpha/r$. In this case, the solution can be expressed in elementary functions \cite{Lyakhovich:2025}.    
 If the precession constant is small, $K<\frac{M^2}{8m}$, the overall centrifugal energy will remain repulsive, and the effective potential $U_{eff}(r)$ will be similar to the usual Kepler problem. With positive energy the motion is unbounded. With negative energy, the motion is bounded, but it is aperiodic unless $K\neq 0$ \cite{Lyakhovich:2025}. The trajectories precess, unlike the usual Kepler case. For $K=0$, we have ordinary Kepler trajectories. For a large value of the precession constant $K>\frac{M^2}{8m}$, the centrifugal force becomes attractive and vanishes at $K=\frac{M^2}{8m}$. For $K\geq\frac{M^2}{8m}$, there is no potential well. For a negative total energy, the particle always falls toward the center. For a positive energy, it either goes to infinity or falls into the center, depending on the sign of the initial velocity $\dot{r}$. 
All unbounded trajectories in this problem are spiral-shaped except for the case with $K=0$. 

We also note that if the constraint $\dot{M}=0$ was included in the Lagrangian with a Lagrange multiplier, the system would require six initial conditions, due to the additional equation for the multiplier. Meanwhile, equations (\ref{C-L-E-M}) require five conditions. This additional degree of freedom, unrelated to the conditional extremum, would contribute to the Noether conserved quantities for the action with multipliers. These quantities, as a result, would not characterize the conditional extremals of the action.

\noindent
We see that although the effective potential energy is similar to the case of an unconditional extremum, an additional constant of motion can qualitatively change the dynamics.

\subsection*{Example 2. Particle in central field with conserved energy.} 
Let us consider the same action of the point particle (\ref{S-particle}), but restrict the admissible class of trajectories by imposing another constraint, conservation of energy,
\begin{equation}\label{E-dot}
\frac{dE}{dt}=0\, ,\quad E=\frac{m\dot{r}^2}{2}+\frac{mr^2\dot{\phi}^2}{2}+U(r) \, .
\end{equation}
Similar to the previously considered condition $\dot{M}=0$, this constraint is a consequence of the Lagrange equations for the problem of the unconditional extremum of the action.
But if this condition is imposed from the very beginning as a constraint, then the admissible class of trajectories will include curves that do not necessarily obey the Lagrangian equations.Therefore, critical lines for the conditional extremum may include non-Lagrangian trajectories. To construct the conditionally Lagrangian equations for the action (\ref{S-particle}) we have at first to find the gauge symmetry transformation of the  constraint (\ref{E-dot}). Applying the general method of the article \cite{Lyakhovich:2008hu}, we find the symmetry 
\begin{equation}\label{GT-E}
\delta_\epsilon r=\dot{r}\epsilon+\frac{r\dot{\phi}}{\kappa}\dot{\epsilon},\quad \delta_\epsilon\phi=\dot{\phi}\epsilon-\frac{\dot{r}}{r\kappa}\dot{\epsilon},
\end{equation}
where
\begin{equation}\label{kappa}
\kappa=-\frac{2\dot{r}\dot{\phi}+r\ddot{\phi}}{\dot{r}}-\frac{2r\dot{\phi}U^{\prime}(r)}{m\dot{r}^2+mr^2\dot{\phi}^2}.
\end{equation}
This leads to the conditionally Lagrangian equations
\begin{equation}\label{CLE-E}
\resizebox{0.9\textwidth}{!}{$\displaystyle
\frac{\delta S}{\delta\epsilon}=-mr\dot{r}\dot{\phi}^2-m\ddot{r}-\frac{dU}{dr}
-mr^2\dot{\phi}\ddot{\phi}-\frac{d}{dt}\left[\frac{1}{\kappa}\left(mr^2\dot{\phi}^3-mr\ddot{r}\dot{\phi}-r\dot{\phi}\frac{dU}{dr}
-\frac{\dot{r}}{r}\frac{dM}{dt}\right)\right]=0\, ;\quad\frac{dE}{dt}=0
$}
\end{equation}
These equations define the critical trajectories for the particle action in the class of lines restricted by the constraint $\frac{dE}{dt}=0$. 
 
It is curious to mention the degenerate case $U(r)\equiv 0$. The Lagrangian in this case coincides with the energy, so the imposed constraint 
means $\dot{L}=0$ along any admissible trajectory. Since the integrand in the action is a constant in the admissible class of lines, the requirement of the conditional extremum should not impose further restrictions on the trajectory, besides the constraint, while the first equation in the system (\ref{CLE-E}) may seem independent from the second one. In fact, it is not. With $U(r)\equiv 0$, equations (\ref{CLE-E}) read
\begin{equation}\label{CLE-E-U-0}
\frac{\delta S}{\delta\epsilon}\equiv -\frac{dE}{dt}+\frac{d}{dt}\left[2E+\frac{r\dot{\phi}}{\dot{r}\kappa}\frac{dE}{dt}\right]=0\, , \qquad  \frac{dE}{dt}=0 \, .
\end{equation}
In this specific case, as we see, the first equation in the system (\ref{CLE-E}) reduces to the differential consequences of $\dot{E}=0$. This does not happen with any non-trivial potential $U(r)$. 

Let us consider the issue about symmetries and conserved quantities for the conditionally Lagrangian system (\ref{CLE-E}). The Lagrangian and the constraint do not explicitly involve time and angle, so the time shift and rotation are obvious symmetries. In the case of the usual Lagrange equations, this would immediately lead to conservation of energy and angular momentum. For a conditionally Lagrangian system, we first have to represent the time translation and rotation as a specialization of gauge symmetry (\ref{GT-E}) and identify the corresponding conditional symmetry generators. Given the generators, we can construct the conserved quantities by the recipe of the previous section.

Choosing the gauge parameter $\epsilon(t)=\omega=const$ in  (\ref{GT-E}), we obviously arrive at time translation, much like in the previous example. This is because the time derivative of the gauge parameter is involved in the transformations (\ref{GT-M}) and (\ref{GT-E}) in a different way, while the terms  without derivatives coincide. Since this is the same transformation as in the previous example, and the Lagrangian is the same, it transforms to a total derivative, as in the previous case (\ref{delta-omega-L-M}). However, this does not mean that the conserved quantity associated with this conditional symmetry is the same. It is different because the general expression for the conserved quantity (\ref{I-A}), (\ref{tilde-I}) includes the gauge generator itself, not just a specialization. Substituting the time shift generator and corresponding gauge generators into (\ref{I-A}), (\ref{tilde-I}) we find the conserved quantity
\begin{equation}\label{barE}
\overline{\! E}=\frac{1}{\kappa}\left(mr^2\dot{\phi}^3-mr\ddot{r}\dot{\phi}-r\dot{\phi}\frac{dU}{dr}+\frac{\dot{r}}{r}\frac{dM}{dt}\right),
\quad M=mr^2\dot{\phi}\, .
\end{equation}
This quantity is conserved due to the homogeneity of time, and it complements, in a sense, already imposed conservation of energy (\ref{E-dot}). We call this quantity the \emph{energy of precession}. 
Using the constraint (\ref{E-dot}), this constant of motion can be reduced to another form, which is more convenient from the perspective of reducing the problem to quadratures,
\begin{equation}\label{barE-2}
\overline{\! E}=\frac{1}{\dot{r}\kappa}(2\dot{r}\dot{\phi}+r\ddot{\phi})(m\dot{r}^2+mr^2\dot{\phi}^2)=
-\frac{dM}{dt}\left[\frac{d}{dt}\left(\frac{M}{m\dot{r}^2+mr^2\dot{\phi}^2}\right)\right]^{-1} \, .
\end{equation}
Now, let us consider another specialization of the gauge symmetry (\ref{GT-E})
\begin{equation}\label{epsilon-omega-E-phi}
\epsilon=\frac{M}{m\dot{r}^2+mr^2\dot{\phi}^2}\omega, \quad \omega=const \, .
\end{equation}
Given the constraint (\ref{E-dot}), the transformation (\ref{GT-E}) with the above gauge parameter reduces to the shift of the angle
\begin{equation}
\delta_\omega r=0, \quad \delta_\omega\phi=\omega \, .
\end{equation}
So it is a rotation. Since the Lagrangian depends only on $\dot{\phi}$ and does not involve $\phi$, it is rotationally invariant without any shift by the total derivative, $\delta_\omega L=0$. Given the gauge symmetry (\ref{GT-E}), and the specialization (\ref{epsilon-omega-E-phi}),  making use of the general formula (\ref{I-A}), we find the conserved quantity $\overline{\! M}$,
\begin{equation}\label{M-ext}
\overline{\! M}= \frac{M}{\Gamma(r)} \, ,\quad \Gamma(r)=\frac{E-U(r)}{E+\frac{\overline{\! E}}{2}-U(r)} \, ,\quad   M=mr^2\dot{\phi}\, .
\end{equation}
This constant of motion could also be found by directly integrating the relation (\ref{barE-2}) for conservation of energy. The integration is obvious as the right hand side of relation (\ref{barE-2}) is the ratio of the total derivatives, while the left hand side is a constant. 
Since this quantity is conserved due to the isotropy, it should be considered as the  angular momentum of the conditionally Lagrangian system (\ref{CLE-E}). 
As we see, $\overline{\! M}$ differs from the usual angular momentum $M$ by the factor $\Gamma$ involving the energy of precession $\overline{\! E}$ which is conserved due to homogeneity of time in this system. Once $\overline{\! E}\rightarrow 0$, we have $\Gamma\rightarrow 1$ and $\overline{\! M}\rightarrow M$.
Solutions with $\overline{\! E}=0$ exist, and they obviously reproduce the usual Lagrangian dynamics where the angular momentum $\overline{\! M}$ reduces to $M$.
 
Much like the previous example, we have now three constants of motion: $E,\,\overline{\! M},\,\overline{\! E}$. 
The first of them is imposed from the outset as a constraint. The second conserved quantity is connected with rotation, 
and the third is connected with time shift.  Substituting the last two conserved quantities into the first, we arrive at the relation
 \begin{equation}\label{E-new}
E=\frac{m\dot{r}^2}{2}+ U_{eff}(r) =const\, , \qquad U_{eff}(r)=\frac{\overline{\! M}{}^2}{2mr^2}\Gamma^2(r)+U(r)\, ,\quad 
\Gamma(r)=\frac{E-U(r)}{E+\frac{\overline{\! E}}{2}-U(r)}\, .
\end{equation}  
As we see, the distinction from the case of the unconditional extremum is in the modification of the centrifugal potential energy, which involves now the additional constant of motion  -- the energy of precession  $\overline{\! E}$ -- much like the previous example, where another extra constant $K$ is involved.   In the limit $\overline{\! E}\rightarrow 0$ the factor $\Gamma \rightarrow 1$ and we arrive at the effective potential energy of the usual problem of central field for the unconditional extremum of the action (\ref{S-particle}). This is similar to the previous example. 
A slight difference from the previous example is that for certain values of the constants of motion (including $\overline{\! E}$) and the shape of U(r), the effective potential can have poles at points other than the center. These poles can define impassable potential barriers or infinite potential wells. 

Relation (\ref{E-new}) can be resolved in quadratures with respect to $r$ in the usual way,
\begin{equation}\label{int-r-E}
  t-t_0=\pm\int\limits_{r_0}^r \frac{dr}{\sqrt{\frac{2}{m}\left(E-U_{eff}(r,\, E,\, \overline{\! M},\, \overline{\! E})\right)}} \, .
\end{equation}
Given conservation of the angular momentum (\ref{M-ext}), and $\dot{r}$ defined by (\ref{E-new}), 
the angular velocity $\dot{\phi}$ can be expressed as a function of $r$. This allows us to find $\phi(r)$ in quadratures for any potential $U(r)$,
\begin{equation}\label{phi-E}
\phi -\phi_0=\pm \int\limits_{r_0}^{r}\,\frac{\overline{\! M}\Gamma(r)}{r^2}\frac{dr}{\sqrt{2m\left(E-U_{eff}(r,\, E,\, \overline{\! M},\, \overline{\! E})\right)}} \, .
\end{equation}
If the precession energy vanishes $\overline{\! E}=0$, corresponding solutions reproduce the ordinary Lagrangian trajectories for the particle in a central potential. 

Unlike the previous example, for the Kepler potential $U=-\alpha/r$ these integrals are not expressed in terms of elementary functions, unless $\overline{\! E}\neq 0$, so studying the geometry of the precessing trajectories becomes a tricky issue. The limit $\overline{\! E}\rightarrow 0$ obviously brings the integrals (\ref{int-r-E}), (\ref{phi-E}) to the known
Kepler solutions.

Let us make a short overall conclusion to the two examples of mechanical models. 

At first let us mention that the method of a Lagrange multipliers in these examples would result in the extra degree of freedom comparing to the problem of conditional extremum.  Conditionally Lagrangian system in both cases includes one equation of the third order, and one more of the second order. If the imposed constraint was included into the action with Lagrange multiplier, the Lagrangian system would include three independent equations of the second order. Obviously, this would require an extra initial condition related to the Lagrange multiplier. 

In both cases, a conservation law from the unconditional extremum problem was imposed as a constraint on the admissible class of trajectories. The corresponding global symmetry of the action turned out to be representable as a specialization of the gauge symmetry of the imposed constraint. Then, applying the recipe of Theorem 1, we arrive at an additional conservation law associated with this symmetry, complementary to the one imposed as a constraint. The conditionally Lagrangian equations remain solvable in quadratures due to this additional constant of motion, though their order increase comparing to the case of the unconditional extremum. The solutions of this problem reproduce the trajectories of the unconditional extremum problem when this additional constant of motion vanishes. Depending on the value of this additional conserved quantity, the solutions can either slightly deviate from the case of the unconditional extremum, or the motion can be qualitatively different. 

\subsection*{Example 3. Proca action with imposed condition of transversality}
Consider the Proca action 
\begin{equation}\label{S-Proca}
S=\int dx^4 (-\frac{1}{4}F^{\mu\nu}F_{\mu\nu}-\frac{m^2}{2}A^\mu A_{\mu}) \, ,
\end{equation}
where $F_{\mu\nu}=\partial_\mu A_\nu-\partial_\nu A_\mu$ and Minkowski metrics is assumed mostly positive $\eta_{\mu\nu}=diag\lbrace-1,1,1,1\rbrace$. \newline
Let us impose transversality condition on the vector field
\begin{equation}\label{div-A}
\partial_\mu A^\mu=0 \, .
\end{equation}
If the admissible class of the fields was not restricted by this equation, it would appear anyway as a differential consequence of the Proca equations. The class of transverse fields is obviously broader then the set of Proca fields. Therefore, there may be conditional critical field configurations for the action that do not necessarily satisfy Proca's equations. To obtain the field equations for this problem, we have to find the gauge symmetry of equation (\ref{div-A}) restricting the class of admissible fields. In principle, finding a complete gauge symmetry of a system of nonlinear non-Lagrangian field equations in any dimension can be a nontrivial task. There is a systematic method for $d=2$ \cite{Lyakhovich:2013qfa}, while we do not know any general procedure of finding infinitesimal gauge symmetry for $d>2$. In specific cases, the gauge symmetry can be known from the outset, or can be found by ad hoc methods. For linear field equations, the problem of finding a gauge symmetry reduces to the Hilbert syzygy problem in any dimension (see, for example, \cite{Piontkovski}), so it can always be solved using the appropriate algorithms of commutative algebra. For equation (\ref{div-A}), the gauge symmetry is obvious,
\begin{equation}\label{GT-Transverse}
\delta_\epsilon A^\mu=\partial_\nu\epsilon^{\nu\mu} \, , \qquad \epsilon^{\mu\nu}=-\epsilon^{\nu\mu} \, ,
\end{equation}
where the gauge parameter $\epsilon^{\mu\nu} (x)$ is an arbitrary antisymmetric tensor field.

Conditionally Lagrangian equations (\ref{Partially-Lagrangian}) for the Proca action in the class of fields restricted by equation (\ref{div-A}) read
\begin{equation}\label{CLE-Proca}
\frac{\delta S}{\delta\epsilon^{\mu\nu}} = (\Box-m^2)F_{\mu\nu}=0 \, , \quad \partial_\mu A^\mu=0 \, .
\end{equation}
\noindent
Now, the issue is to find the energy-momentum tensor for the system above. Following the general prescription of Section 3, 
we have to find the specialization of the  gauge transformation (\ref{GT-Transverse}) such that reproduce the Poincar\'e shift of the vector field,
\begin{equation}\label{Poincare-shift}
  \delta_\omega A_\mu=\omega^\nu\partial_\nu A_\mu\, , \quad \omega^\nu=const \, .
\end{equation}
Let us consider the following specialization of the gauge parameter
\begin{equation}\label{spec-Proca}
  \epsilon_\omega^{\mu\nu}=\omega^\mu A^\nu -\omega^\nu A^\mu \, .
\end{equation}
With this special parameter, the general gauge transformation reads
\begin{equation}\label{delta-omega-A}
  \tilde{\delta}_\omega A^\mu =\omega^\nu\partial_\nu A^\mu -\omega^\mu\partial_\nu A^\nu \, .
\end{equation}
In general, this transformation differs from the Poincare shift, but coincides with the translation (\ref{Poincare-shift}) for transverse fields satisfying (\ref{div-A}). So, the specialization of gauge transformation (\ref{GT-Transverse}), (\ref{spec-Proca}) is a conditional symmetry indeed.
Note that the Proca Lagrangian, being explicitly independent of $x$, transforms into a total divergence under the Poincar\'e shift (\ref{Poincare-shift})
\begin{equation}\label{delta-omega-Proca}
  \tilde{\delta}_\omega L(A,\partial A) =\omega^\mu\partial_\nu\left(\partial^\lambda A^\nu\partial_\mu A_\lambda\right)+\omega^\nu\partial_\nu L=\omega^\mu\partial _\nu\left(\partial^\lambda A^\nu\partial_\mu A_\lambda+\delta_\mu^\nu L\right) \, .
\end{equation}
Substituting the gauge generator (\ref{GT-Transverse}), generator of conditional symmetry (\ref{spec-Proca}) and the total divergence (\ref{delta-omega-Proca}) into the general formula for the conserved current (\ref{I-A}), (\ref{tilde-I}), we arrive at the conserved tensor
\begin{equation}\label{T-E-M-transverse}
T^\mu_\nu=\delta^\mu_\nu(A_\lambda\Box A^\lambda-\frac{1}{2}m^2A_\lambda A^\lambda+\frac{1}{2}\partial_\lambda A_\rho\partial^\lambda A^\rho)-A^\mu(\Box-m^2)A_\nu-\partial_\nu A_\lambda\partial^\mu A^\lambda \, .
\end{equation}
It can be seen that this tensor is indeed conserved by virtue of the conditionally Lagrangian equations (\ref{CLE-Proca}),
\begin{equation}\label{divT-Proca}
  \partial_\mu T^\mu_\nu = -A^\mu(\Box -m^2)F_{\mu\nu}-\left((\Box-m^2)A_\nu\right) \partial_\mu A^\mu = 0 \, .
\end{equation}
Conservation of this tensor is connected with homogeneity of Minkowski space, so it should be understood as the energy-momentum of the conditionally Lagrangian system (\ref{CLE-Proca}).

The solutions of Proca equations obey a more relaxed conditionally Lagrangian system (\ref{CLE-Proca}). Let us consider the reduction of the energy-momentum tensor (\ref{T-E-M-transverse}) to the fields subject to Proca equations:
\begin{equation}\label{T-Proca}
T^\mu_\nu \vphantom{\begin{subarray}{l} \Box A^\lambda \mapsto m^2 A^\lambda \\ \partial_\lambda A^\lambda=0 \end{subarray}}_{\Big| \begin{subarray}{l} \Box A^\lambda\, \mapsto\, m^2 A^\lambda \\ \partial_\lambda A^\lambda\,=\,0 \end{subarray}}
= \frac{1}{2}\delta^\mu_\nu(\partial_\lambda A_\rho\partial^\lambda A^\rho+m^2A_\lambda A^\lambda)-\partial_\nu A_\lambda\partial^\mu A^\lambda \, .
\end{equation}
This is a canonical energy-momentum tensor for the Proca Lagrangian. 

Let us also mention that equation (\ref{div-A}), being considered by itself, is a pure gauge system. The  transfor\-ma\-ti\-ons (\ref{GT-Transverse}) gauge out all local degrees of freedom if the dynamics is governed solely by the transversality equation. In the cases of this special type, the action with Lagrange multipliers (\ref{S-lambda}) should lead to the system which is equivalent to the conditionally Lagrangian equations (\ref{Partially-Lagrangian}). Hence, the energy momentum for the Proca action with Lagrange multiplier to $\partial_\mu A^\mu$ should reduce on shell to (\ref{T-E-M-transverse}). This fact can be easily verified. As we see, the Lagrange multiplier method, in cases where it does not introduce additional degrees of freedom, indeed leads to conserved quantities that are  equivalent on the mass shell to the currents constructed by the recipe of Theorem 1 in terms of the original fields.   

One can note that the field theory based on the conditional extremum of the Proca action in the class of transverse fields allows for the conservation of the energy-momentum tensor, which includes the tensor of a massive spin-one field. Without going into detail, we note that the system (\ref{T-E-M-transverse}) describes four local degrees of freedom\footnote{ This can be counted in various ways. For example, one can use the method of article \cite{Piontkovski}.} in $d=4$. In addition to the obvious three degrees of freedom associated with the massive vector, there is also a massless scalar that contributes to the energy-momentum tensor. 

The appearance of additional degrees of freedom compared to the unconditional extremum problem when imposing constraints that are differential equations is a fairly natural phenomenon if the gauge transformations of the imposed equations include derivatives of the parameters. This is evident from the very form of equations (\ref{Partially-Lagrangian}), the order of which in this case increases compared to the Lagrangian equations. 

\section{Concluding remarks}
Noether's theorem in it's original form connects symmetries of the action with the currents conserved by virtue of corresponding Lagrangian equations.
In this paper, we extend this connection to the class of theories whose field equations follow from a conditional principle of least action, where the class of admissible field configurations is constrained by imposing a system of differential equations. We do not introduce Lagrange multipliers or any other auxiliary fields and construct conserved quantities that include only the original degrees of freedom. The key ingredient of our construction is a self-contained system of field equations (\ref{Partially-Lagrangian}) that define the conditional extremals of the action.  
Equations (\ref{T}) that constrain the admissible fields are a part of this system. The global symmetry of the action, if it is a specialization of the gauge symmetry of these constraints, turns out to be connected with currents that are conserved by virtue of the conditionally Lagrangian equations. This connection is a one-to-one correspondence, as established by the two theorems of Section 3. These theorems can be viewed as an extension of Noether's first theorem and its converse to the case of a conditional extremum of the action.

We conclude with some general remarks on the conditional extremum problem itself, beyond the connection between global symmetries and conserved quantities.
The idea of restricting the class of field configurations admissible for action minimization by imposing differential equations may at first glance seem to implicitly threaten the locality of the theory, since explicit reduction to solutions may be non-local. Including constraints with Lagrange multipliers into the action makes the problem explicitly local, but the multipliers can introduce additional degrees of freedom that violate, generally speaking, the equivalence with the original conditional extremum problem. Consequently, in field theory and mathematical physics more broadly, the conditional extremum problem does not gain much interest, except for cases where the Lagrange multiplier method might be applicable. In the paper \cite{Lyakhovich:2025}, it is proposed to construct a complete system of differential equations that determine the conditional extremals of the action using the infinitesimal gauge symmetry of the equations that restrict the admissible class of fields. These equations for conditional extremals are explicitly local and they turn out to admit Hamiltonian formulation \cite{Lyakhovich:2025}. Hence, these classical theories can be quantized, at least at the level of deformation quantization. The corresponding Poisson bracket is degenerate, in general. The structure of the bracket is defined by the gauge algebra of the equations constraining the trajectories. The same infinitesimal gauge symmetry plays a key role in the connection between symmetries and conservation laws in the problem of the conditional extremum of the action. In conventional Lagrangian field theories, gauge structures play an important role in a variety of ways \cite{Henneaux:1992ig}, including issues of quantization, conserved charges, and the inclusion of interactions. We see that in the much less studied class of theories based on the conditional action principle, in addition to the gauge symmetry of the theory as a whole, the gauge symmetry of the equations constraining the admissible field configurations also plays a significant and distinct role.We hope that the revelation of this role, as well as the further development of general methods for describing gauge theories based on the principle of conditional least action, will lead to the emergence of new meaningful models of this class.

 We would like to mention a few more general considerations about how so far underestimated type of conditional extremum problems could expand the possibilities for posing various problems in field theory. Differential equations that restrict the class of admissible fields select a submanifold in the configuration space of fields. These equations are contained as a special subsystem in the full system of field equations (\ref{Partially-Lagrangian}) that determines the conditional extremals of the action. If this subsystem is considered as such, separately from the full system of conditional Lagrangian equations, and it describes certain degrees of freedom that are not gauged out, then these degrees of freedom will have, in a certain sense, a different status. They describe the geometry of the configuration space of admissible fields, rather than the dynamics in this space, which is determined by the minimization of the action. The dynamics of these degrees of freedom are fixed in the strong sense, and therefore, with a suitable quantization scheme, they should not fluctuate. Roughly speaking, the directions admissible for quantum field fluctuations are determined by gauge generators of transformations (\ref{GT}). This idea is implicitly confirmed by the structure of the Poisson bracket (see \cite{Lyakhovich:2025}) for the conditionally Lagrangian equations. This implicit separation of degrees of freedom into "variational" and "non-variational" ones, occurring in the conditional extremum problem, creates new options for describing various interactions. As a somewhat speculative example, we can mention the following idea. The equations of conformal gravity can be imposed as constraints restricting the class of admissible metric  configurations, and a non-conformal action is chosen, for example, the Einstein-Hilbert action. Then, only the local scale mode of the metric will be quantized, while the modes of conformal gravity will evolve classically. In this case, no explicit separation of these degrees of freedom will be required. We will not speculate further about other possible ideas of this kind, but we hope that posing problems on a conditional extremum will prove useful in one way or another.

\vspace{0.2 cm}
 
\textbf{Acknowledgements.} 
The work is supported by the research project FSWM-2025-0007 of the Ministry
of Science and Higher Education of Russian Federation.

\noindent
The authors thank M.A.~Grigoriev and A.A.~Sharapov for the valuable discussions. 

\noindent
We are also thankful to V.A.~Abakumova for useful remarks on the examples of Section 4.

\end{document}